# Quantum Annealing based Hybrid Strategies for Real Time Route Optimization

Sushil Mario[a], Pavan Teja Pothamsetti[a], Louie Antony Thalakottor[a], Trisha Vishwanath[a], Sanjay H.A[c], Anees Ahmed[b], Salvatore Sinno[b], Shruthi Thuravakkath[b], Sinthuja M[a]

[a]Department of Information Science and Engineering, RIT, Bangalore 560054
[b]Unisys India Private Limited, Bangalore
[c]BMS Institute of Technology and Management, Bangalore

Email: { sushil.mario@gmail.com, 1ms19is088@gmail.com, louieantony@gmail.com, trisha.vishwanath15@gmail.com, sanjay.ha@bmsit.in, anees.ahmed@in.unisys.com, salvatore.sinno@gb.unisys.com, shruthi.thuravakkath@unisys.com, sinthuja@msrit.edu }

## Abstract

One of the most well-known problems in transportation and logistics is the Capacitated Vehicle Routing Problem (CVRP). It involves optimizing a set of truck routes to service a set of customers, subject to limits on truck capacity, to reduce travel costs. The biggest challenge faced whilst attempting to solve the issue is that the time complexity of the issue grows exponentially with the number of customers and trucks, rendering it virtually intractable to traditional computers and algorithms. In this paper, we propose a method to circumvent this limitation, employing quantum computers to aid classical computers in solving problems faster while reducing complexity. To obtain our results, we employ two algorithms: Hybrid Two Step (H2S) and Hybrid Three Step (H3S). Both algorithms involve two phases: clustering and routing. It has been observed that both algorithms produce promising results, both in terms of solution time and solution cost.

## Abbreviations

**CVRP** (Capacitated Vehicle Routing Problem), **FCM** (Fuzzy C Means), **H3S** (Hybrid Three Step), **H2S** (Hybrid Two Step), **QA** (Quantum Annealing), **QUBO** (Quadratic Unconstrained Binary Optimization), **QPU** (Quantum Processing Unit), **TSP** (Traveling Salesman Problem).

## 1. Introduction

The transportation and logistics sector are a vital aspect of any consumer driven economy since it enables the transfer of goods and services from the centers of production to those of consumption. The logistics and transportation systems must be efficient in terms of both cost and time to ensure the seamless operation of businesses and industries and the satisfaction of customers[1]. In reality, however, the current state of transportation and logistics is marked by various obstacles, including rising fuel costs, environmental issues, traffic congestion, and suboptimal routing and scheduling of trucks[2]. This results in increased operational costs, extended delivery times and lower profit margins for businesses. Possessing a robust vehicle routing algorithm is the need of the hour for businesses that want to succeed and thrive in a competitive market.

Several optimization techniques have been employed to handle real-time route optimization issues. However, they are restricted by the problem's magnitude, intricacy and frequently do not deliver optimal solutions within a reasonable timeframe [3].

Rietsche et al. [4], in their paper, state, "A whole new ecosystem around quantum computing technology itself is emerging already, provoking questions around the change of (1) business models and process innovation, (2) challenges for IT organizations, or (3) sourcing from start-ups, full-stack providers such as Google, IBM, Microsoft, Alibaba or individual development". With such new concepts and ideas emerging, we believe that logistics and transportation planning will be positively impacted.

The most famous theoretical problem in the domain is that of the travelling salesman[5], which has proven to be quite challenging to solve, especially for large instances. Given a collection of cities and the distance of travel between each pair, the traveling salesman's problem (TSP) aims to find the shortest route to traverse, such that each city is visited exactly once before returning to the starting point[6]. Obtaining optimal delivery routes by solving the TSP can enhance resource usages, such as delivery trucks, reducing empty miles and increasing operational efficiency[7].

The Capacitated Vehicle Routing Problem (CVRP), a variant of the TSP, is a combinatorial optimization problem that is similarly relevant to the logistics and transportation industry [8]. The problem states that given a collection of n customers, each with a known demand and location, and a fleet of m trucks, each with a given capacity, the objective is to identify a set of truck routes that serves all customers while minimizing total distance traveled. Ibrahim Obaid [9] investigates the CVRP as a complex optimization problem that presents considerable difficulties that need to be solved efficiently. Their paper outlines that this is because the CVRP is known to be NP-hard, a class of problems known to be so computationally challenging that they cannot be solved in polynomial time. Quantum Annealing (QA) is a relatively new optimization technique that has been proposed as a potential solution method for the CVRP in yielding promising results[10].

Quantum annealing is an adaptation to the well-known meta-heuristic simulated annealing, which employs temperature fluctuations to guide a solution down an energy landscape [11]. QA exploits principles of quantum mechanics in addition to the natural tendency of any system to exist in the least possible energy state, allowing for a more efficient traversal of the landscape. As will be outlined in this study, QA has shown promise in solving route optimization issues more effectively than its counterparts in classical computing[12].

This work aims to investigate the current status of research on CVRP, TSP, and QA, recognize the prevalent issues, and explore innovative methods to tackle them in the quest to attain optimal logistics and transportation routes. In this capacity, we compare results obtained from two solution methods: Hybrid Two Step and Hybrid Three Step algorithm.

The paper is organized as follows: Section II provides an overview of numerous themes and an essential background on key terms that is necessary for a more thorough understanding of the paper. A literature review that provides a summary of the current research in the field is then included in Section III. In Section IV, we describe the concepts and ideas that underpin the H3S and H2S algorithms. Section V presents the implementation of the proposed methodology of the algorithms presented in Section IV. Following that, we present the strategies' outcomes in Section VI. In Section VII, we provide a brief summary of the entire research project and outline a potential future research path.

## 2. Background

In this section, we aim to acquaint readers with the core concepts and terminologies facilitating our proposed solutions and the dataset employed to test their efficiency.

VRPlib, the dataset we have selected, consists of numerous problem instances for the CVRP compiled from real-world data in Brazil[13]. The parameters in VRPlib provide information about the customers, including their location in the form of (x, y) coordinates the quantity of goods that each customer requires to be delivered (demand), and the maximum amount of goods that can be carried by each truck (capacity).

Optimization problems involve finding the best solution from a set of possible solutions that are bounded by various constraints and measured by an objective function, as defined by decision variables[14]. Optimisation aims to find an optimal and feasible set of values for the decision variables, one that minimizes or maximizes the value of the objective function while simultaneously satisfying all constraints. The problem can be constructed in various ways, depending on how the objective and constraints are represented. Quadratic Unconstrained Binary Optimization (QUBO) is the representation identified to be most compatible with QA, and hence is our choice of problem formulation[15].

The problem is first represented in standard mathematical notation, with the decision variables, objective and constraints clearly defined. Conversion to QUBO format involves first replacing discrete and continuous-valued variables with binary equivalents[16]. The CVRP lends itself naturally to the task, as the primary decision is whether to allocate a truck to a customer (1) or not (0). What follows is unconstraining, where inequality constraints are transformed into equations using Lagrange parameters. Finally, quadratization reduces polynomial expressions to equivalent quadratics by introducing new binary variables and mappings that relate them to the previous variables.

The solution space to an optimization problem can be represented as an energy landscape in n-dimensional space. Each particle (point) in the landscape represents a single solution. Peaks are associated with high energy values, while valleys are associated with low energy values. The Hamiltonian is a mathematical operator used in physics to represent the total energy contained within a physical system[17]. It is like every such system to seek and settle in a stable state with the lowest possible energy[18]. Thus, a lower value of the Hamiltonian indicates a more optimal solution than a higher value. Therefore, the objective function of an optimization problem can be represented in the form of a Hamiltonian, and the problem becomes one of minimization.

QA is a technique that employs quantum mechanical principles to aid in locating the global minimum of an objective function. This is achieved by using small fluctuations in magnetic fields to guide particles through the energy landscape. Variations in the field produce different values of the variables and, thus, different solutions. A property of quantum particles that enables them to perform the traversal of the landscape more efficiently is quantum tunnelling, which allows them to tunnel through an energy barrier, represented by a peak, instead of ascending and then descending it [19]. In addition, the particles can exist in a combination of different states simultaneously, thus permitting consideration of several parallel solutions instead of in sequence[20].

The purpose of QA is to identify the lowest value of the objective, represented by a Hamiltonian in QUBO formulation. It accomplishes this by employing quantum processing units (QPU' 's) to generate a quantum state representing all potential solutions to the QUBO problem. This quantum state is then "annealed" until it settles into the lowest energy state, representing the best solution to the problem[21]. The quantum system is started in a simple, known state and then allowed to evolve according to a Hamiltonian that encapsulates the objective of the optimization problem. The Hamiltonian is gradually altered over time by annealing until the system achieves the ground state, which represents the ideal solution to the optimization issue. One of the primary advantages of quantum annealing over classical techniques is the possibility for exponential speedup for specific optimization problems[22], such as those that can be translated onto the Ising model, a variant of QUBO that utilizes spin variables. However, fruitful application of QA must overcome various hurdles, including the requirement for high-quality qubits shielded from external noise and the capacity to sustain quantum coherence for an extended period, the absence of which leads to a devolution of the quantum state.

A feasible solution to an optimization problem satisfies all the constraints associated with the problem. In order to accommodate them in the energy landscape, constraints can be transformed into penalties and included as part of the objective function[23]. A penalty introduces a large positive value if its condition is met, thereby increasing the Hamiltonian and rendering the solution less optimal. Constraints can be either in the form of an equation or an inequality.

Lagrange parameters are tunable values traditionally introduced in operations research when solving optimization problems when converting inequalities in constraints to equations using slack and surplus variables[24]. In our use case, when introducing constraints as penalties to the objective function, Lagrange parameters serve as weights to prioritize the satisfaction of a constraint. High values will tend to orient the solution towards strict feasibility at the expense of optimality in terms of high solution cost and vice versa.

For a solution to the CVRP to be optimal, it must consider the spatial arrangement of customers, or risk increased transportation costs due to inefficient routes[25]. Also, solving an unaltered CVRP using current, state-of-the-art quantum annealers is a time-consuming and energy-inefficient process. Meeting complex requirements for constraint satisfaction takes away from the quest for an optimal solution with a low cost for the objective function. Scalability to larger instances is another issue, stemming from QPUs being limited in their number of qubits to maintain coherence[26]. This can be partially alleviated by tasking additional qubits to counteract noise, but this confers an overhead in time and cost. However, these shortcomings can be addressed by employing a hybrid approach of both classical and quantum solvers, working in tandem with their strengths[27].

The hybrid approach involves dividing the CVRP into two sub-problems, namely capacitated clustering and routing. Clustering is typically performed using a classical algorithm such as K-Means, adapted to include capacity and demand satisfaction. It serves to group individual customers into m clusters (as many as there are trucks) based on proximity and capacity considerations. Due to its computationally intensive nature that involves repeated mathematical calculations, clustering lends itself well to computation on a classical computer[28]. On the other hand, routing between customers in a cluster benefits greatly from evaluating multiple potential solutions in parallel, an apt use case for a quantum computer[29]. The size of the solution space is reduced by clustering, aiding the QA algorithm in finding an optimal solution in an efficient manner. This approach is appealing as it addresses the limitations of both classical and quantum computers, using each to supplement the ' 'other's weaknesses. The CVRP is thus decomposed into m instances of the TSP, with each truck assigned to a cluster serving as its "salesman'.

In this study, we propose two hybrid approaches, hybrid two-step (H2S) and hybrid three-step (H3S), that both employ the Fuzzy C-means (FCM) algorithm for clustering and QA for routing.

## 3. Literature Survey

The main bottleneck associated with real-world applications involving the CVRP is the time it takes to compute solutions for large instances. By breaking the problem down into smaller instances, Francesco et al. offered a novel way to reduce size. In order to make the original challenge more manageable, their study describes an effective method of clustering that considers the situation's limitations. This results in a smaller truck routing problem with fewer possible outcomes. One of the two strategies in this paper, Hybrid Three Step (H3S), was largely inspired by their technique, Constrained Clustering CVRP (CC-CVRP). Other CVRP clustering studies frequently assign vehicle clusters based on proximity to local optima.

In contrast to earlier studies, CC-CVRP solves a high-level CVRP by treating clusters as compressed nodes, then decompresses the nodes back into individual customers and performs routing between them. CC-CVRP enables routing between a variety of compressed clusters to enable a variety of tours, as opposed to simply assigning clusters to trucks based on proximity and getting stuck in a local optimum. The simulation results show a large decrease in processing times without a corresponding decline in solution quality, which lowers the problem's complexity.

An alternate version of the TSP, known as quadratic unconstrained binary optimization (QUBO), is shown in the publication by Jain [30]. It outlines their strategy for fixing the issue on an Ising Hamiltonian-based quantum annealer, namely the 5,000 qubit Advantage 1.1 quantum annealer from D-Wave, and highlights their methods. The experiment's findings showed that the quantum annealer only found problems with eight or fewer customers tractable and performed poorly in terms of time and accuracy compared to a classical solver.

In their study, Borcinova [31] described a flow formulation like CVRP and proposed additional limitations addressing the elimination of sub-tours, i.e., cycles that do not pass through the depot and whose cardinality grows exponentially with the number of clients. They contrasted this strategy with another formulation of mixed linear programming with polynomial cardinality of subtour removal requirements. The findings revealed that it is impractical to solve these constraints at once and that it thus spent a significant amount of CPU time to obtain optimal solutions since the number of inequalities of the sub tour elimination constraints in the earlier formulation rises exponentially with the number of nodes. The latter version is more computationally efficient since it contains polynomial size constraints or $O(n^2)$ constraints. This paper added value and gave us insights into using subtour elimination constraints more effectively.

The entire energy of a quantum system is expressed mathematically using the Hamiltonian in quantum computing. The Hamiltonian operator, represented by the letter H, is frequently expressed in terms of the quantum mechanical states of a system. It is specifically used to describe a quantum system's dynamics across time. The energy associated with each term in the Hamiltonian, which typically takes the form of a group of terms that each describe a different component of the quantum system, is expressed. For instance, the Hamiltonian in a system of two interacting qubits can have components that reflect the energy of each qubit and terms that represent the interaction between the two qubits.

The Travelling Salesman Problem with Time Windows (TSPTW), a variation of the travelling salesman problem (TSP), was thoroughly examined by Ozlem et al.[32] in their work. The idea behind TSPTW is to locate the least expensive tour that stops in each city within a predetermined time frame, i.e., between the tour's earliest start and due times. The study also covered edge-based and integer linear programming (ILP) formulations, the researchers introduced QUBO and higher order binary optimization formulations. The formulations were assessed using examples from three, four, and five cities. It was evident from comparing the experimental results that the ILP formulation showed more potential because situations involving four cities had a higher probability of spotting a sample encoding the perfect route. They asserted that the different energy landscapes may have caused the disparity in performance between the two formulations that each one produced.

Quantum annealing is a computing type that uses quantum-mechanical phenomena to solve optimization problems. The purpose of optimization problems, which can occur in many fields of science, engineering, and finance, is to select the best option from a large pool of alternatives. The basic idea underlying quantum annealing is to find the system's lowest energy state using a quantum system, such as a quantum annealer or a quantum simulator. To do this, the Hamiltonian is gradually changed from an initial state that is simple to prepare to a final state that encodes the solution to the optimization problem. Quantum annealing's key advantage over conventional optimization techniques is its ability to explore the solution space more thoroughly by exploiting quantum tunnelling and entanglement. Quantum tunnelling allows the system to "tunnel" beyond energy barriers that would be insurmountable in traditional systems, while entanglement allows the system to study multiple choices simultaneously.

The rigorous idea of time, which asserts that if one city is far from the others, travelling there will take more time than travelling to the other towns, is a problem with most formulations, according to Hirotaka et al. [33]. The researchers suggested a unique technique for formulating CVRP as QUBO to overcome this restriction. The formula contains a timeline depicting each truck's evolution over time. This made it possible to meet a variety of time constraints. Additionally, limits on capacity were implemented, allowing capacitated volumes to vary according to the locations from which trucks arrived. The proposed formulation successfully runs in small-size QUBO systems, but real-world applications need at least more than 2000 logical qubits, according to the findings of testing it on the D-Wave 2000Q.

The term "quantum hybrid" describes blending classical and quantum computing. While quantum computers can handle some problems much more quickly than conventional computers, they also have drawbacks, including high error rates and the requirement for enormous numbers of qubits to execute practical computations. This is why quantum hybrid systems are used. By employing a hybrid strategy, quantum computers can concentrate on the computationally difficult portions of the problem, while classical computers can handle activities like error correction, optimization, and data pretreatment. This makes it possible to combine the benefits of classical and quantum computing for improved accuracy and efficiency.

The specific formulation of the optimization problem has been a persistent problem in addressing the CVRP on the quantum annealers. It must be translated into a quadratic unconstrained binary optimization (QUBO) problem to achieve this. A hybrid method is presented by Feld et al. [34] to address the CVRP issue. The approach uses a two-phase heuristic that combines clustering and routing. Three models were proposed: the Quantum 2 model, the Quantum 1 model, and a hybrid model to carry out the two-phase heuristic. In the hybrid method, the researchers solved the routing problem using a quantum algorithm after doing the clustering using a conventional technique. They proposed that the issue may be broken down into more manageable subproblems, allowing for a sequential resolution of the partitioned issue. The study's findings indicated no discernible advantages regarding computing time or solution quality. This article served as the foundation for our strategy for implementing a quantum hybrid formulation.

The objective of clustering is to arrange a collection of items so that they are more similar to one another than those in other groupings. These units can alternatively be thought of as several groupings. A technique that created a soft-clustered vehicle routing problem with fewer decision variables was shown by Alesiani et al. [35]. They explained how their formulation, which decreased the computational cost of the original CVRP solution, and how the calculated solution could be converted back into the original spaces. The outcomes showed enhanced, nearly real-time solutions with less optimality gaps.

# 4. System Design

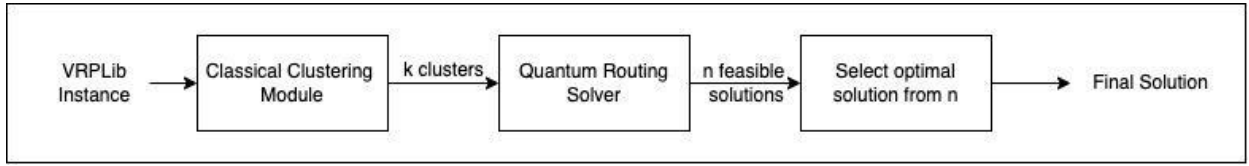

**Fig. 1** Overview of the Proposed System

The solutions we propose in the form of H2S and H3S can be visualized using a pipeline, as detailed in the flowchart in Fig. 1, which consists of two main steps.
   a. Clustering
   b. Routing

During clustering, the customer nodes are assigned to clusters on the basis of both proximity to the cluster centroid and sufficient remaining cluster capacity to accommodate the customer's demand. For H2S, only a single cluster is assigned to a truck, so the capacity of each cluster is equal to that of each truck. For H3S, multiple clusters can be assigned to each truck, so the capacity of each cluster is a fraction of that of each truck.

Next is the solution of the routing problem. For H2S, each cluster is treated as a problem instance for the TSP and solved to obtain the optimal route covering all customers. The solution pipeline for H3S introduces an intermediate step between clustering and the solution of TSP, which is the solution of a CVRP. A CVRP is formulated using the clusters obtained in the previous clustering step. The cluster centroids are treated as compressed nodes, essentially pseudo-customers in their own right. The solution of this CVRP results in an assignment of trucks to nodes. Each truck's assigned clusters are then expanded into individual customer nodes, which serve as the input for the TSP solver.

The difference between the steps that constitute H2S and those for H3S is further highlighted in Table 1.

| Component | H2S | H3S |
|---|---|---|
| Classical FCM Clustering | Yes | Yes |
| Quantum CVRP Solver | No | Yes |
| Quantum TSP Solver | Yes | Yes |

**Table 1.** Difference between H2S and H3S

## 4.1. Classical Clustering Module

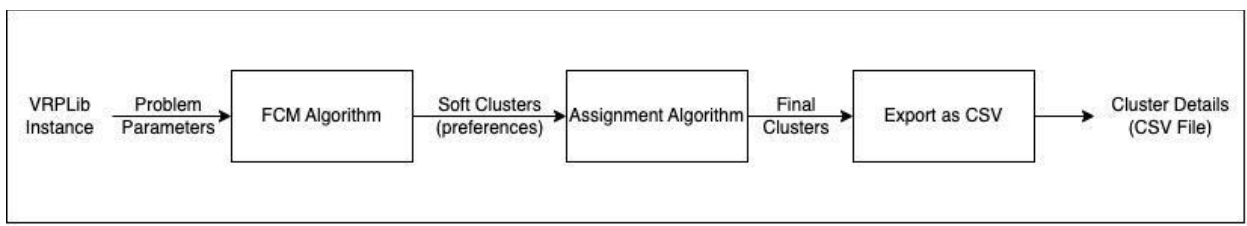

**Fig. 2** Overview of Classical Clustering Module

Fig. 2 represents the first step in our solution pipeline, wherein the instance parameters, including the number and location of customers and trucks, are passed in as input and a list of membership values of each customer associated with each cluster is received as output.

The algorithm chosen for the approaches presented in this paper is Fuzzy-C-Means clustering. It was chosen primarily for its ability to produce soft clusters. Clustering can be broadly divided into two categories, namely hard and soft clustering. Hard clustering produces a rigid assignment of nodes to clusters, whereas soft clustering generates a list of membership values, denoting the affinity expressed by a node toward joining a specific cluster. Soft clustering allows for final assignment to be performed at a later stage, at which point the demand of a customer and the capacity of a truck can be considered.

The algorithm begins by randomising the cluster centroids and the membership matrix, which defines the mapping of membership values from nodes to centroids. The cluster centroids and degree of membership are then updated repeatedly until convergence. The cluster centroids are updated based on the weighted average of the data points, where the weights represent the degrees of membership. The membership matrix is revised after each iteration based on the separation between each data point and each cluster centroid. The working of the algorithm can be explained as a sequence of the following five steps: -

Step 1: Random initialization of data points

For $p$ clusters, each data point lies in all the $p$ clusters with some membership value, which can be assumed to be anything in the initial state.

Step 2: Calculating the centroids

The formula for calculating centroid $V$ is

$$V_{ij} = (\sum_{1}^{n}(\gamma_{ik}^m * x_k)) / \sum_{1}^{n} \gamma_{ik}^m \qquad (1)$$

Where $\gamma$ is fuzzy membership value of the data point, $m$ is the fuzziness parameter (generally taken as 2), and $x_k$ is the data point. $i$ is the cluster number, $j$ is the coordinate (abscissa or ordinate of the cluster centroid), $k$ is the data point number, $n$ is the total number of data points.

Step 3: Calculate the distance of each point from the centroid

Calculate the Euclidean distance $d$ of each point from each centroid given by the formula

$$d = \sqrt{(x_2 - x_1)^2 + (y_2 - y_1)^2} \qquad (2)$$

Where $(x_1, x_2)$ and $(y_1, y_2)$ represent the two points.

Step 4: Updating membership values

$$\gamma_{ik} = [\sum_{1}^{n}(d_{ki}^2/d_{kj}^2)^{1/m-1}]^{-1} \qquad (3)$$

Here, $i$ is the cluster number, $j$ is the coordinate (abscissa or ordinate of the cluster centroid), $k$ is the data point number, $n$ is the total number of data points.

Step 5:

Repeat the steps (2-4) until the constant values of the member values are obtained or the value is less than the tolerance value. The tolerance value is the small value up to which the difference between the values of two consecutive updates is accepted. Its value can be assigned depending on the number of data points and clusters.

Step 6: Defuzzification of the membership values

Assign the nodes to the clusters with the help of the membership values.

To better suit our particular use case, we have made the following additions to the original FCM algorithm: incorporating additional constraints and modifying the objective function to better reflect the problem requirements.

- Create as many copies of the depot as there are clusters, before executing the FCM algorithm.
- Employ the elbow method using sum of intra-cluster distances to select an optimal value of c, the number of clusters

The algorithm for classical clustering serves to output a set of centroids, not medoids; the difference is that centroids may or may not be a member of the set of points, whereas medoids are required to be. Thus, the denotation of the depot as centroid/medoid of all clusters is not mandatory, as that would detract from the assignment algorithm's ability to perform an effective assignment. However, cluster centroids located at a considerable distance from the depot would diminish solution quality of the routing algorithm by increasing travel cost. In our implementation of classical clustering, we strove to maintain a delicate balance between the two extremes. The approach decided on was to incorporate multiple instances of the depot (by making the requisite number of copies), as many as there are clusters. This serves to provide additional weightage to the depot and thus skew the clusters so that the centroids are located closer to the depot than the actual center of the resulting clusters.

## 4.2. Assignment Algorithm

The input to this step consists of a set of nodes $n$ with a corresponding set of membership values $M = \{m_{ij} : 0 \leq m_{ij} \leq 1; m_{ij} \in R, \ i \in N, \ j \in C\}$ (R being the set of real numbers), representing affinities of each node $n$ towards each cluster centroid $c_i \in C$.

Each cluster centroid is initialized with a capacity as given by the equation:

$$capacity_{cluster} = \frac{capacity_{truck}}{\frac{n_{clusters}}{n_{trucks}}} \quad (4)$$

The algorithm is composed of two modules. Module 1 handles the actual assignment of nodes to preferred clusters, while Module 2 runs Module 1 to completion.

**Module 1** - Assignment based on preference

- List of node preferences - results of classical clustering
- Find the cluster centroid with the highest preference for each node
    - Pop the highest preference
- Construct a list for each cluster centroid with only nodes whose highest preference is that centroid.
- Sort the list in decreasing order of membership value
- Add nodes to each cluster's assignment array until capacity is met

**Module 2** - Repeat module 1 until all nodes are assigned

We begin the assignment algorithm by associating each node with its most preferred cluster centroid, $c_j$. Then, for each cluster centroid $c_j$, a list of nodes $N_j$ that have centroid $c_j$ as their most preferred centroid is generated. In the next step, these lists are sorted in descending order of membership value. Finally, for each centroid, each node in its corresponding list of nodes $N_j$ is iteratively assigned to the ' 'centroid's cluster, the demand corresponding to the node being subtracted from the cluster's capacity. Once the current cluster has insufficient capacity to accommodate the next node in its list, the next centroid $c_{j+1}$ is considered, and so on, for all $c_j \in C$.

The algorithm is run in a loop, each iteration eliminating the most preferred centroid from the list of available candidates for each unassigned node. The loop terminates once all nodes have been assigned.

## 4.3. Optimization Problem Formulation using Hamiltonians

The solution space to an optimization problem can be represented as an energy landscape in n-dimensional space. Peaks are associated with high energy values, while valleys are associated with low energy values. The Hamiltonian is a mathematical operator used in physics to represent the total energy contained within a physical system. It is like every such system to seek and settle in a stable state with the lowest possible energy. Quantum Annealing seeks to exploit this natural tendency to guide the solution down the energy ' 'landscape's peaks towards the valleys, thus minimizing the ' 'system's Hamiltonian. A lower value of the Hamiltonian indicates a more optimal solution than a higher value. Therefore, the objective function of an optimization problem can be represented in the form of a Hamiltonian, and the problem becomes one of minimization.

A feasible solution to an optimization problem satisfies all the constraints associated with the problem. In order to accommodate them in the energy landscape, constraints can be transformed instead into penalties and included in the objective function. A penalty introduces a large positive value into the objective function if the constraint is violated, thereby rendering the solution less optimal. Constraints can be either in the form of an equation or an inequality. Thus, The penalties must be formulated creatively and carefully to model the constraints accurately.

## 4.4. QA on CVRP and TSP

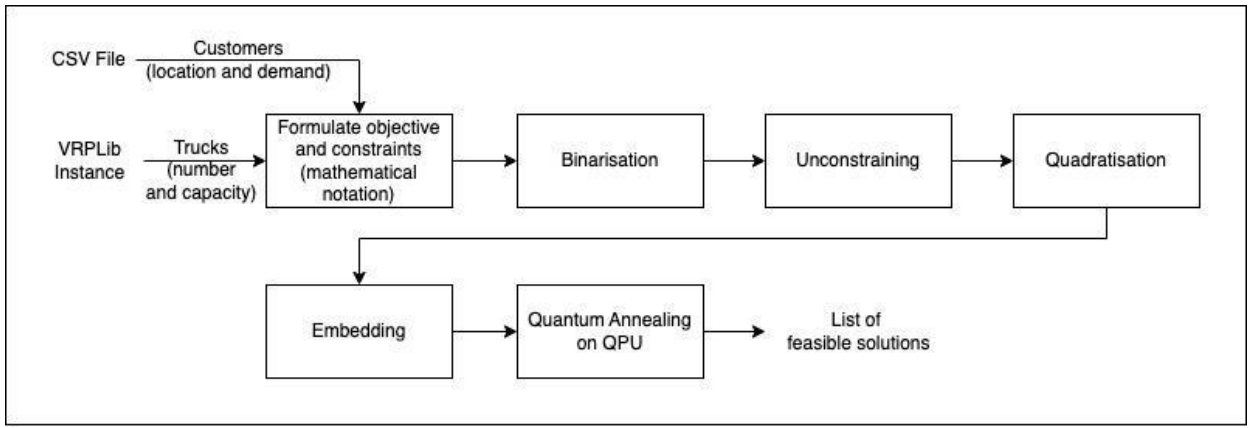

**Fig. 3** Overview of Quantum Routing Solver

### 4.4.1. QA

Quantum annealing is a quantum computing methodology that uses the laws of quantum mechanics to solve optimization problems following the flow presented in Fig. 3. It is especially well suited to handling difficult combinatorial optimization problems, requiring simultaneous evaluation of several potential solutions.

Combinatorial optimization issues for QA are typically modelled mathematically using QUBO, or Quadratic Unconstrained Binary optimization [36]. This is due to the fact that it gives users a means to communicate optimization issues in a way that quantum annealing software and hardware, such as D-Wave quantum annealers, are primed to comprehend.

A quadratic function with binary variables with values of 0 or 1 must be minimized in a QUBO problem.

QUBO equations essentially consist of two components:

- Objective Function
- Constraints

The general form of a QUBO problem is:

$$minimum\ f(x) = x^T Q x \qquad (5)$$

where:

- $x$ is a vector of binary variables representing the constraints considered.

$$x_i \in \{0, 1\}$$

- $Q$ is a symmetric matrix representing the objective function
- $x^T$ denotes the transpose of the vector $x$

### 4.4.2. CVRP

From the previous step, i.e., clustering, we obtain a set of n clusters with one or more customer nodes assigned to them. Each cluster has a centroid located at the arithmetic mean of coordinates of all assigned nodes. Also associated with each cluster is an aggregate demand, representative of the sum of demands of all assigned customer nodes. The cluster centroids are now considered to be compressed pseudo-customer nodes. The location of the pseudo-customers corresponds to that of the centroids, and its demand corresponds to the aggregate demand of the ' 'centroid's cluster. This is modelled as a CVRP instance to be solved by the QA CVRP solver. This step occurs only for H3S, not H2S.

In order to solve the Capacitated Vehicle Routing Problem using a QUBO formulation, we have derived the following objective function and accompanying constraints. These accurately represent the problem's requirements in terms of optimality and feasibility, respectively.

The decision variables of the optimization problem are used to represent the path traversed by each truck. They can be defined as follows: -

$$x_{r,i,j} = \begin{cases} 1 & ; \ truck \ r \ traverses \ edge \ (i,j) \\ 0 & ; \ truck \ r \ does \ not \ traverse \ edge \ (i,j) \end{cases}$$

$$s_{i,r} = \begin{cases} 1 & ; \ node \ i \ is \ visited \ by \ truck \ r \\ 0 & ; node \ i \ is \ not \ visited \ by \ truck \ r \end{cases}$$

**Objective function**

$$min(\sum_{r=0}^{p-1} \sum_{i=0}^{n-1} \sum_{j=0}^{n-1} c_{i,j} * x_{r,i,j}) \qquad (6)$$

$c_{i,j}$ is the cost for the edge (i,j)

**Constraint 1 - Single Visit**

Each node is visited only once in the entire solution (including all trucks).

**Mathematical Formulation**

$$\sum_{i=0, i \neq j}^{n-1} \sum_{r=0}^{p-1} x_{r,i,j} = 1, \forall \ j \in N \qquad (7)$$

**Hamiltonian Equation**

$$H = \sum_{j=1}^{n-1}(\sum_{i=0,i\neq j}^{n-1}\sum_{r=0}^{p-1} x_{r,i,j} - 1)^2 \qquad (8)$$

For any destination node j, excluding node 0 (the depot, which will be visited once by every truck, not just one), a summation is set up so that $x_{r,i,j} = 1$ for only a single value of $i$, across all trucks $r$. If this holds for all nodes j, $H$ will have the minimum possible value of 0.

**Constraint 2 - Depot First**

Each truck leaves the depot only once.

**Mathematical Formulation**

$$\sum_{r=0}^{p-1}\sum_{j=0}^{n-1} x_{r,0,j} = 1 \qquad (9)$$

**Hamiltonian Equation**

$$H = \sum_{r=0}^{p-1}(\sum_{j=1}^{n-1} x_{r,0,j} - 1)^2 \qquad (10)$$

Here, the source node i is set to the constant value of 0, denoting the depot. For every truck, $x_{r,0,j} = 1$ for only a single value of j ensures $H$ is at its minimum value of 0. This signifies that only one edge with the depot as the source is traversed and thus the truck only leaves the depot once.

**Constraint 3 - Route Validity**

If truck t visits a node n, truck t must next leave node n

**Mathematical Formulation**

$$\sum_{j=0}^{n-1}\sum_{r=0}^{p-1}\sum_{i=0}^{n-1}(x_{r,j,i} - \sum_{k=0}^{n-1} x_{r,j,k}) = 0 \ \forall \ i \neq j \neq k \qquad (11)$$

**Hamiltonian Equation**

$$\sum_{j=0}^{n-1}\sum_{r=0}^{p-1}(\sum_{i=0,i\neq j}^{n-1} x_{r,i,j} - \sum_{i=0,i\neq j}^{n-1} x_{r,j,i})^2 \qquad (12)$$

This constraint ensures that if a truck $r$ enters node $j$ from some node $i$ ($x_{r,i,j} = 1$) it must also leave node $j$ to some other node $i$ ($x_{r,j,i} = 1$). However, this leaves the door open for a loop from $i$ to $j$ and then back to $i$. This constraint is used in conjunction with constraint 5 (sequence of steps) to account for this potentially undesirable occurrence.

**Constraint 4 - Capacity**

Capacity of a truck is not exceeded

**Mathematical Formulation**

$$\sum_{j=0}^{n-1}\sum_{i=0}^{n-1} D_j * x_{r,i,j} \leq Q, \ \forall \ j \neq i, \ r \ in \ [0,p) \qquad (13)$$

$D_j$ is the demand of the node j

$Q$ is capacity of each truck

**Hamiltonian Equation**

$$H = \sum_{r=0}^{p-1}[\sum_{j=1}^{n-1}\sum_{i=0}^{n-1}(D_j * x_{r,i,j}) - Q] \qquad (14)$$

For each truck $r$, the demand of node $j$ is considered in the summation if it is the distance of an edge included in the truck's route, i.e., when $x_{r,i,j} = 1$. This thus represents the total demand a truck aims to fulfill. An inequality is set up by subtracting the capacity of the truck from this, ensuring that the capacity is greater than total demand. If not, a large positive value is added to the objective function by means of the Lagrange parameter associated with this constraint.

**Constraint 5 - Sequence of Steps**

Eliminate any and all sub-tours, for each truck. A sub-tour is a circuit in a graph that does not include the depot

**Mathematical Formulation**

$$\sum_{i=1}^{n-1} \sum_{j=1, j \neq i}^{n-1} \sum_{r=0}^{p-1} s[j][k] - (s[i][k] + 1) + B * (1 - x[k][i][j]) \qquad (15)$$

**Hamiltonian Equation**

$$H = \sum_{i=1}^{n-1} \sum_{j,j=1, j \neq i}^{n-1} \sum_{r=0}^{p-1} -[s_{j,r} - (s_{i,r} + 1) + B * (1 - x_{r,i,j})] \qquad (16)$$

In this constraint formulation, a second binary decision variable $s$ is introduced, denoting the step in the sequence of routes that a node is traversed by a particular truck. A mapping between $x$ and $s$ is induced in the second part of the equation, where $B$ is a tunable parameter. If $x_{i,j,k} = 1$, a negative value of H (desirable) is contingent on node $j$ being visited after node $i$. If $x_{i,j,k} = 0$, the sufficiently large positive value of B that comes into play ensures that the value of H is negative.

### 4.4.3. TSP

The same objective function and constraints, save for the demand constraint, used for the Capacitated Vehicle Routing Problem were employed for the Travelling Salesman Problem. This works because the Capacitated Vehicle Routing Problem is a generalization of the Travelling Salesman Problem. There is just one truck that acts as a stand-in for the salesman, and the demand constraint is removed as it has already been accounted for in the clustering phase. This step is common to both the H2S and H3S approaches.

## 5. Experimental Setup and Results

To test our approaches, the problem instances that belong to class A described in Augerat et al. (1995) have been used. The publicly accessible version of these test instances and their globally optimal solutions may be found at http://vrp.galgos.inf.puc-rio.br.

The code for clustering and routing was written in Python. The former was executed in an interactive Python notebook (ipynb) on Google Colab, and the latter on Leap, an IDE provided by DWave, a company that manufactures quantum annealing hardware. Clustering was performing with the aid of the skfuzzy Python library, and quantum annealing (including setting up QUBO equations and submission to the annealer for processing) were performed using the dimod library that is part of 'DWave's Ocean API.

To solve the CVRP problem in this work, the methodology uses two distinct approaches: H3S and H2S. Their outcomes are shown in this section. In the tables below, we have provided our findings for both approaches using several instances from the VRPlib dataset. The best-known solution for the situation is listed in the table along with our findings for each strategy. For a clearer understanding of the overall outcomes, we have also computed and provided the optimality gap between the best-known solution and our findings for each of the instances.

| Instance | Best Known Solution | Proven Optimal | H3S | H2S |
|---|---|---|---|---|
| A-n32-k5 | 784 | Yes | 855 | 873 |
| A–n33-k5 | 661 | Yes | 712 | 695 |
| A-n34-k5 | 778 | Yes | 874 | 832 |
| A-n36-k5 | 799 | Yes | 896 | 938 |
| A-n37-k5 | 669 | Yes | 766 | 763 |

**Table 2.** H3S and H2S solutions

We first implemented H3S for 5 instances and then compared it to the results of H2S. From table 1, one can note that the H3S solutions are significantly lesser than the H2S solutions for instances A–n33-k5, A-n33-k6, A-n34-k5, A-n37-k5, while H2S outperforms H3S in instances A-n32-k5 and A-n36-k5. There was no feasible solution found for the H2S instance of A-n33-k6. It was discovered that, on average, the computational time for the quantum annealer is 15 minutes to solve the problem while executing each instance of the VRPlib dataset.

| Instance | Customer Distribution | Depot Location | Approach to Return Lower Cost |
|---|---|---|---|
| A-n32-k5 | Scattered | Corner | H3S |
| A–n33-k5 | Clustered | Center | H2S |
| A-n34-k5 | Scattered | Center | H2S |
| A-n36-k5 | Clustered | Corner | H3S |
| A-n37-k5 | Scattered | Center | H2S |

**Table 3.** Analysis of instances and best approaches

Following the testing of the two approaches on the instances, we proceeded to undertake a cursory analysis of the distribution of customers and location of the depot, manually assigning classes for each, as shown in Table 3. We define customer distribution as 'clustered' if a clear separation of customers into groups is conspicuous, and as '''scattered' if not. Location of the depot is denoted as '''corner' if the depot is present along the periphery of the solution space spanned by the instance, and '''center' if it is instead surrounded by customers. A preliminary observation was that H3S seemed to perform better on instances with depot location as '''corner' and H2S fared better on instances with depot location as ''center'. Further testing will be needed to confirm if this pattern is indeed statistically significant.

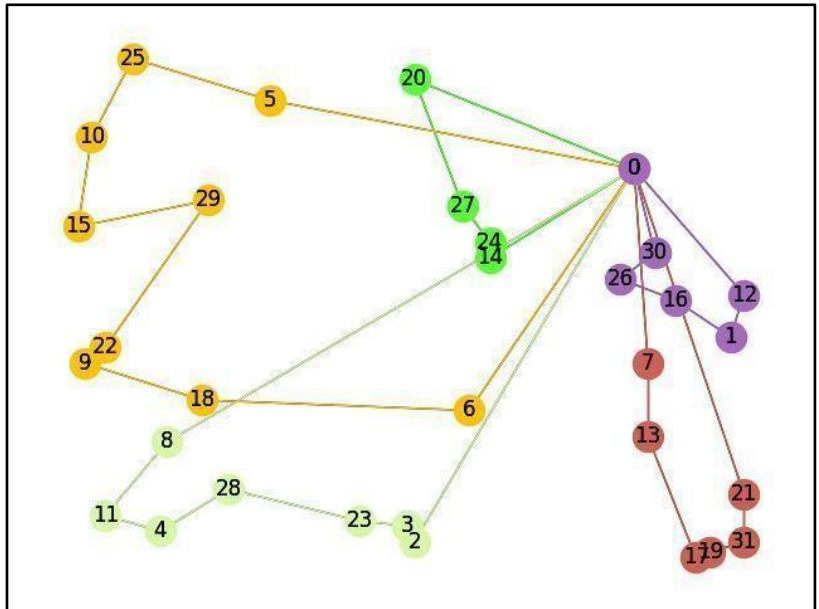

**Fig.6** Optimal routes for instance A-n32-k5 (H3S)

     Fig.6 depicts 32 customers that have been assigned to different clusters and are serviced by five trucks. Node' '0' represents the depot from which each truck forms closed loops, leaving and returning only once. As illustrated in the Fig.4, each color-coded loop portrays an individual cluster that is optimal. However, there are certain cases where a node might have been wrongly assigned to a cluster, decreasing the optimal solution. For example, node 6 has been assigned to the yellow cluster but should ideally be included in the light green cluster.

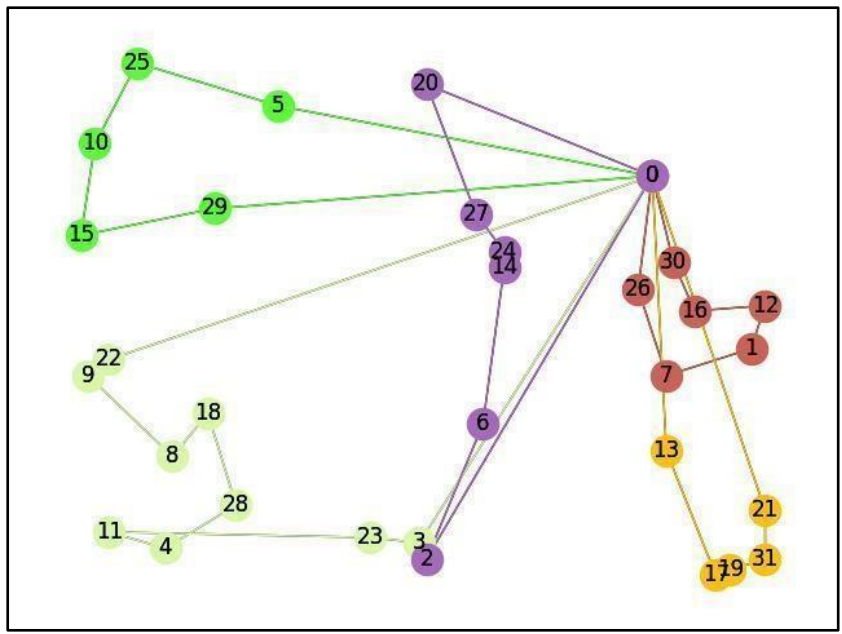

**Fig.7** Optimal routes for instance A-n32-k5 (H2S)

Fig. 7 shows the graph that was produced after the instance A-n32-k5 was run through the H2S approach. We have provided a comparable instance for each algorithm to make comparing the two algorithms easier. Fig. 5 shows that the route shown in the purple color coding appears to cross two additional routes, which causes the overall cost to rise.

## 6. Conclusion

This study evaluated the performance of two classical clustering and QA-based approaches to solving the Capacitated Vehicle Routing Problem (CVRP) on a publicly available dataset. The two approaches were the Hybrid Three Step technique and the Hybrid Two Step (H2S) approach. The initial stage of the H3S technique required clustering over the customer data points in the instance. The centroid of the clusters is derived once the customers are assigned to their respective clusters. CVRP was then performed on the centroids, and TSP on the CVRP results. In the H2S technique, the H2S algorithm is used to cluster customer data points, and then the CVRP algorithm is applied to the clusters to find the optimal least cost routes, solving a TSP. The findings of experimentation on different instances are detailed, and a comparison between the approaches is drawn.

As future scope, we foresee great potential in further investigating sophisticated methods of clustering that use learning based algorithms to reproduce the tight configurations seen in the heretofore best classical solutions. This, when paired with concomitant research into additional constraints to avoid common pitfalls and advancements in quantum annealing technology to support more qubits to handle larger problem sizes, can facilitate the deployment of classical-quantum hybrid algorithms in real world scenarios to assess reliability and feasibility of solving the CVRP in real-time.